\documentclass{acmsiggraph}                

\usepackage{epsfig, xspace, color, setspace, booktabs}
\title{N-Body Simulations on GPUs}
\onlineid{80}
\author{
 Erich Elsen \hspace{0.2in}
 V. Vishal\hspace{0.2in}
 Mike Houston \\
 Vijay Pande\hspace{0.2in}
 Pat Hanrahan\hspace{0.2in}
 Eric Darve
}
\affiliation{Stanford University}

\acmformat{print}

\contactname{Mike Houston}
\contactaddress{Gates Building, Room 381 \\
                Stanford University \\
                Stanford, CA  94305}
\contactphone{(650) 723-0618}
\contactfax{(650) 723-0033}
\contactemail{mhouston@graphics.stanford.edu}

\estpages{9}

\newcommand{\nvidia}{NVIDIA}

\newcommand{\nbody}{$N$-body }
\newcommand{\floatfour}{\texttt{float4}}
\newcommand{\fah}{Folding@Home}

\RequirePackage[normalem]{ulem}
\RequirePackage{color}\definecolor{RED}{rgb}{1,0,0}\definecolor{BLUE}{rgb}{0,0,1}

\newsavebox{\savepar}

\newenvironment{codepar}{\scriptsize}
{}

\begin{document}

\maketitle

\begin{abstract}

Commercial graphics processors (GPUs) have high compute capacity at
very low cost, which makes them attractive for general purpose
scientific computing. In this paper we show how
graphics processors can be used for \nbody simulations to obtain
improvements in performance over current generation CPUs. We have
developed a highly optimized algorithm for performing the $O(N^2)$
force calculations that constitute the major part of stellar and
molecular dynamics simulations. In some of the calculations, we
achieve sustained performance of nearly 100 GFlops on an ATI
X1900XTX.  The performance on
GPUs is comparable to specialized processors such as
GRAPE-6A and MDGRAPE-3, but at a fraction of the cost.
Furthermore, the wide availability of GPUs has significant
implications for cluster computing and distributed computing efforts
like \fah.

\end{abstract}


\keywordlist

\copyrightspace

\section{Introduction}

The classical \nbody problem consists of obtaining the time evolution 
of a system of $N$ mass particles interacting according to a given force law.
The problem arises in several contexts, ranging from molecular scale
calculations in structural biology to stellar scale research in astrophysics.

Molecular dynamics (MD) has been successfully used to understand how certain
proteins fold and function, which have been outstanding questions in biology for
over three decades~\cite{MD-review,gomez_2004}.
Exciting new developments in MD methods offer hope that such
calculations will play a significant role in future drug research~\cite{FEM-Hide}. In stellar
dynamics where experimental observations are hard, if not impossible, theoretical
calculations may often be the only way to understand the formation and evolution
of galaxies.

Analytic solutions to the equations of motion
for more than 2 particles or complicated force functions are intractable which forces one to resort to computer
simulations. A typical simulation consists of a force evaluation step,
where the force law and the current configuration of the system are
used to the compute the forces on each particle, and an update step,
where the dynamical equations (usually Newton's laws) are numerically
stepped forward in time using the computed forces. The updated
configuration is then reused to calculate forces for the next time
step and the cycle is repeated as many times as desired.

The simplest force models are pairwise additive, that is the force of
interaction between two particles is independent of all the other
particles, and the individual forces on a particle add linearly. The
force calculation for such models is of complexity $O(N^2)$. Since typical
studies involve a large number of particles ($10^3$ to $10^6$) and the desired
number of integration steps is usually very large ($10^6$ to $10^{15}$), the
computational requirements often limit both the problem size as well as the
simulation time and consequently, the useful information that may be obtained
from such simulations. Numerous methods have been developed to deal with these
issues.  For molecular simulations, it is common to reduce the number of 
particles by treating the solvent molecules as a continuum. In stellar
simulations, one uses individual time stepping or tree algorithms to minimize the number of force
calculations. Despite such algorithmic approximations and optimizations, the
computational capabilities of current hardware remain a limiting
factor.

Typically \nbody simulations utilize neighborlists, tree methods or other
algorithms to reduce the order of the force calculations. Previous work
~\cite{BCATS} demonstrated a GPU implementation of a neighbor list
based method to compute non-bonded forces. However, since the GPU so far outperformed
the CPU, the neighborlist creation quickly became a limiting factor. Building
the neighborlist on the GPU is extremely difficult due to the lack of specific
abilities (namely indirected output) and research on computing 
the neighborlist on the GPU is still in progress. Other simplistic simulations
that do not need neighborlist updates have been implemented by others~\cite{JCP_MD}.
However, for small N, we find we can do an $O(N^2)$ calculation significantly 
faster on the GPU than an $O(N)$ method using the CPU (or even with a combination
of the GPU and CPU). This has direct applicability to biological simulations
that use continuum models for the solvent. We note also that in many of the reduced order methods
such as tree based schemes, at some stage an $O(N^2)$ calculation is performed
on a subsystem of the particles, so our method can be used to improve the 
performance of such methods as well.  When using GRAPE accelerator cards
 for tree based algorithms, the host processor takes care of building the tree
and the accelerator cards are used to speed up the force calculation step; GPUs
could be used in a similar way in place of the GRAPE accelerator boards.

Using the methods described
below, we are able to accelerate the force calculation on GPUs over
25 times compared to highly optimized SSE code running on an Intel Pentium 4.
This performance is in the range of the specially designed
GRAPE-6A~\cite{fukushige-2005} and MDGRAPE-3~\cite{MDGRAPE} processors, 
but uses a commodity processor at a much better performance/cost ratio. 

\section {Algorithm}
\label{sec:algorithm}

General purpose CPUs are designed for a wide variety of applications
and take limited advantage of the inherent parallelism in many
calculations.  Improving performance in the past has relied on
increasing clock speeds and the size of high speed cache memories.
Programming a CPU for high performance scientific applications
involves careful data layout to utilize the cache optimally and
careful scheduling of instructions.

In contrast, graphics processors are designed for intrinsically parallel
operations, such as shading pixels, where the computations on one pixel are
completely independent of another. GPUs are an example of streaming processors,
which use explicit data parallelism to provide high compute performance and hide
memory latency. Data is expressed as \emph{streams} and data parallel operations
are expressed as \emph{kernels}.  Kernels can be thought of as functions that
transform each element of an input stream into a corresponding element of an
output stream. When expressed this way, the kernel function can be applied to
multiple elements of the input stream in parallel. Instead of blocking data to
fit caches, the data is \emph{streamed} into the compute units. Since streaming
fetches are predetermined, data can be fetched in parallel with computation. We
describe below how the \nbody force calculation can be mapped to streaming
architectures.

In its simplest form the \nbody force calculation can be described by the
following pseudo-code:
\begin{codepar}
\begin{verbatim}
for i = 1 to N
    force[i] = 0
    ri = coordinates[i]
    for j = 1 to N
        rj = coordinates[j]
        force[i] = force[i] + force_function( ri, rj )
    end
end
\end{verbatim}
\end{codepar}
Since all coordinates are fixed during the force calculation, the force
computation can be parallelized for the different values of i. In terms of
streams and kernels, this can be expressed as follows:


\begin{codepar}
\begin{verbatim}
stream coordinates;
stream forces;

kernel kforce( ri ) 
    force = 0
    for j = 1 to N
        rj = coordinates[j]
        force  = force + force_function( ri, rj )
    end
    return force
end kernel

forces = kforce( coordinates )
\end{verbatim}
\end{codepar}
The kernel \texttt{kforce} is applied to each element of the stream
\texttt{coordinates} to produce an element of the \texttt{forces} stream. Note
that the kernel can perform an indexed fetch from the \texttt{coordinates}
stream inside the j-loop. An out-of-order indexed fetch can be slow, since in
general, there is no way to prefetch the data. However in this case the indexed
accesses are sequential. Moreover, the j-loop is executed simultaneously for
many i-elements; even with minimal caching, \texttt{rj} can be reused for many
$N$ i-elements without fetching from memory thus the performance of this
algorithm would be expected to be high. The implementation of this
algorithm on GPUs and GPU-specific performance optimizations are described in
the following section.

There is however one caveat in using a streaming model. Newton's Third law
states that the force on particle $i$ due to particle $j$ is the negative of the
force on particle $j$ due to particle $i$. CPU implementations use this fact to
halve the number of force calculations. However, in the streaming model, the
kernel has no ability to write an out-of-sequence element (scatter), so
\texttt{forces[j]} can not be updated while summing over the j-loop to calculate
\texttt{forces[i]}. This effectively doubles the number of computations that
must be done on the GPU compared to a CPU.

Several commonly used force functions were implemented to measure and compare
performance. For stellar dynamics, depending on the integration scheme being
used, one may need to compute just the forces, or the forces as well as the time
derivative of the forces (jerk). We have designated the corresponding kernels \emph{GA}
(Gravitational Acceleration) and \emph{GAJ} (Gravitational Acceleration and
Jerk). In molecular dynamics, it is not practical to use $O(N^2)$ approaches
when the solvent is treated explicitly, so we restrict ourselves to continuum
solvent models. In such models, the quantum interaction of non-bonded atoms is
given by a Lennard-Jones function and the electrostatic interaction is given by
Coulomb's Law suitably modified to account for the solvent. The
\emph{LJC(constant)} kernel calculates the Coulomb force with a constant
dielectric, while the \emph{LJC(linear)} and \emph{LJC(sigmoidal)} kernels use
distance dependent dielectrics. The equations used for each kernel as well as
the arithmetic complexity of the calculation are shown in Table~\ref{table:kernels}.

\section{Implementation and Optimization on GPUs}
\subsection{Brook}
\label{sec:brook}

\begin{table*}
\small  
\centering
\scriptsize
\begin{tabular}{lccccccccc}
\toprule
       &         & Flops   &        & Input   & Inner   & BW        & Useful & Giga	&System\\
Kernel & Formula & per     & Unroll & (bytes) & Loop    & (GB/s)    & GFLOPS & Intrxns  &Size\\
       &         & Intrxn. &        &         & Insns.  &           &        & per sec.&\\
\midrule

\addlinespace[5pt]

\begin{minipage}{0.75in}{Gravity\\(accel)}\end{minipage}         & $\frac{m_j}{(r_{ij}^2+\epsilon^2)^{3/2}} \mathbf{r}_{ij}$ 
                        & 19 & 4$\times$4 & 64 & 125 & 19.9 & 94.3 & 4.97 & 65,536\\

\addlinespace[5pt]
\hline
\addlinespace[5pt]

\begin{minipage}{0.75in}{Gravity\\(accel \& jerk)}\end{minipage} & \begin{minipage}{1.9in}{ \begin{center} $\frac{m_j}{(r_{ij}^2+\epsilon^2)^{3/2}} \mathbf{r}_{ij}$ \\ $m_j\Big[ \frac{\mathbf{v}_{ij}}{(r_{ij}^2+\epsilon^2)^{3/2}} - 3 \frac{ (\mathbf{r}_{ij} \cdot \mathbf{v}_{ij})\mathbf{r}_{ij}}{(r_{ij}^2+\epsilon^2)^{5/2}} \Big]$ \end{center}}\end{minipage}
                        & 42 & 1$\times$4 & 128 & 104 & 40.6 & 53.5& 1.27& 65,536\\

\addlinespace[5pt]
\hline
\addlinespace[5pt]

\begin{minipage}{0.75in}{LJC\\(constant)}\end{minipage}	        & $\frac{q_iq_j}{\epsilon r_{ij}^3}\mathbf{r}_{ij} + \epsilon_{ij} \left[ \left(\frac{\sigma_{ij}}{r_{ij}} \right) ^6 - \left( \frac{\sigma_{ij}}{r_{ij}}\right)^{12}\right]$ 
                        & 30 & 2$\times$4 & 104 & 109& 33.6   & 77.6  &  2.59   & 4096 \\

\addlinespace[5pt]
\hline
\addlinespace[5pt]

\begin{minipage}{0.75in}{LJC\\(linear)}\end{minipage}            & $\frac{q_i q_j}{r_{ij}^4}\mathbf{r}_{ij} + \epsilon_{ij}\left[\left( \frac{\sigma_{ij}}{r_{ij}} \right) ^6 - \left( \frac{\sigma_{ij}}{r_{ij}} \right)^{12}\right]$            
                        & 30 & 2$\times$4 & 104 & 107 & 34.5   & 79.5  &  2.65  & 4096 \\

\addlinespace[5pt]
\hline
\addlinespace[5pt]

\begin{minipage}{0.75in}{LJC\\(sigmoidal)}\end{minipage}         & \begin{minipage}{1.9in}{ \begin{center}$\frac{q_i q_j}{\zeta(r_{ij})r_{ij}^3}\mathbf{r}_{ij} + \epsilon_{ij}\left[ \left( \frac{\sigma_{ij}}{r_{ij}} \right) ^6 - \left( \frac{\sigma_{ij}}{r_{ij}} \right)^{12}\right]$ $\zeta(r)=e^{(\alpha r^3 + \beta r^2 + \gamma + \delta)}$\end{center}}\end{minipage}
						& 43 & 2$\times$4 & 104 & 138 & 27.3 & 90.3    & 2.10& 4096 \\

\bottomrule

\end{tabular}
\normalsize
\caption{Values for the maximum performance of each kernel on the X1900XTX.  The instructions are counted as the number of pixel shader assembly
arithmetic instructions in the inner loop. Intrxn = Interaction; Insns = Instructions; BW = Bandwidth.
} 
\label{table:kernels}
\end{table*}

BrookGPU~\cite{BROOKGPU} is a C-like high-level language that can be
used to program GPUs as streaming processors. Streams are stored as
textures and kernels are implemented as fragment programs. The
BrookGPU run-time library can utilize a number of graphics interfaces;
for this work we used the Microsoft DirectX 9.0c API and the Pixel
Shader 3.0 specification~\cite{PS30}.  DirectX~\cite{DX9} provides a
vendor-independent abstraction of hardware features. In the Pixel
Shader 3.0 specification, the shader has access to 32 general purpose,
4-component, single precision floating point (\floatfour) registers,
16 \floatfour\ input textures, 4 \floatfour\ render targets (output
streams) and 32 \floatfour\ constant registers. A shader consists of a
number of assembly-like instructions. Current GPUs have a maximum
static program length of 512 (ATI) or 1024 (\nvidia) instructions.
There is a loop limit of 255 iterations of a loop body, but loops can
be nested to increase the total numbers of iterations. \nvidia\ is limited to
65,535 dynamic instructions and ATI can support an unlimited number.
The BrookGPU compiler translates kernels into a high level shader
language like CG or HLSL, which is then compiled into pixel shader
assembly by an appropriate shader compiler like Microsoft's fxc or
\nvidia's cgc. The graphics driver finally maps the Pixel Shader
assembly code into hardware instructions as appropriate to the
architecture.

\subsection{Precision}
\label{sec:Precision}
Recent graphics boards have 32-bit floating point
arithmetic. Consequently we have done all the calculations in single
precision.  Whether or not this is sufficiently accurate for the answers being
sought from the simulation is often a subject of much debate and the authors do
not intend to settle it here.  We are of the opinion that in many
cases, though certainly not all, single precision is enough to obtain
useful results.  Furthermore, if double precision is necessary, it is usually
not required throughout the calculation, but rather only in a select few
instances.  For reference, GRAPE-6~\cite{Grape-precision} performs the
accumulation of accelerations, subtraction of position vectors and update of
positions in 64-bit \emph{fixed point} arithmetic with everything else in either
36, 32 or 29 bit floating point precision.  It is quite common to do the entire
force calculation in single precision for molecular simulations while using
double precision for some operations in the update step. If and where necessary,
the appropriate precision could be emulated on graphics
boards~\cite{GoStTu05double}.  The impact on performance would depend on where
and how often it would be necessary to do calculations in double precision.

\subsection{General Optimization}
\label{sec:General Optimization}
\begin{figure}
	\centering
	\includegraphics[width=2.5in]{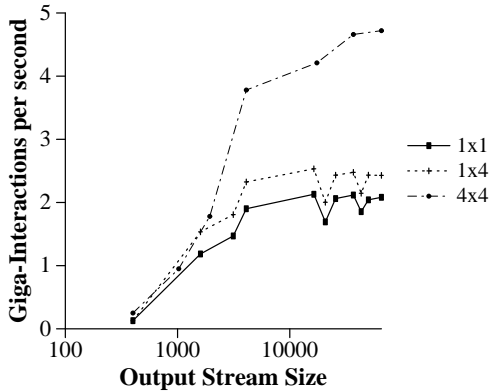}
	\caption{GA Kernel with varying amounts of unrolling}
	\label{fig:all}
\end{figure}

The algorithm was implemented for several force models. For simplicity, in the following discussion,
we only talk about the GA kernel, which corresponds to the
gravitational attraction between two mass particles, given by
\begin{equation} 
\mathbf{a_i} = - G \sum_{i \not= j} \frac{m_j}{({r}_{ij}^2+\epsilon^2)^{3/2}} \mathbf{r}_{ij} 
\end{equation} where $\mathbf{a_i}$ is the
acceleration on particle $i$, $G$ is a constant (often normalized to one), $m_j$ is the mass of
particle $j$, and $\mathbf{r}_{ij}$ is the vector displacement between
particles $i$ and $j$.  The performance of the kernel for various input sizes are shown in
Figure \ref{fig:all}.  

The algorithm outlined in Section~\ref{sec:algorithm} was implemented
in BrookGPU and targeted for the ATI X1900XTX.  Even this naive
implementation performs very well, achieving over 40 GFlops, but its
performance can be improved.  This kernel executes 48
Giga-instructions/sec and has a memory bandwidth of 33 GB/sec.  Using
information from GPUBench~\cite{GPUBENCH}, we expect the X1900XTX to
be able to execute approximately 30-50 Giga-instruction/sec (it
depends heavily on the pipelining of commands) and have a cache memory
bandwidth of 41GB/sec. The nature of the algorithm is such that almost
all the memory reads will be from the cache since all the pixels being
rendered at a given time will be accessing the same j-particle. Thus
this kernel is limited by the rate at which the GPU can issue
instructions (compute bound).

To achieve higher performance, we used the standard technique of loop unrolling.
This naive implementation is designated as a 1$\times$1 kernel because it
is not unrolled in either i or j.  The convention followed hereafter when
designating the amount of unrolling will be that A$\times$B means i unrolled A times
and j unrolled B times. The second GA kernel (1$\times$4) which was written unrolled
the j-loop four times, enabling the use of the 4-way SIMD instructions on the
GPU.  This reduces instructions that must be issued by around a factor of 3. (We
cannot reduce instructions by a factor of 4 because some Pixel Shader
instructions are scalar).  The performance for this kernel is shown in Figure~\ref{fig:all}.  It achieves a modest speedup compared to the previous one, and
we have now switched from being compute bound to bandwidth bound (35
Giga-Instructions/sec and $\approx$40GB/sec).

Further reducing bandwidth usage is somewhat more difficult.  It involves
using the multiple render targets (MRT) capability of recent GPUs
which is abstracted as multiple output streams by BrookGPU.  By
reading in 4 i-particles into each kernel invocation and outputting the force on
each into a separate output stream, we reduce by a factor of four the
size of each output stream compared with original.  This reduces input
bandwidth requirements to one quarter of original bandwidth because each
j-particle is only read by one-quarter as many fragments.  To make
this more clear, we show the pseudo-code for this kernel below.  This
kernel is designated as a 4$\times$4 kernel.

\begin{codepar}
\begin{verbatim}

stream coordinates;
stream index = range( 1 to N skip 4 );
stream forces1, forces2, forces4, forces4;

kernel kforce4x4( i )
    force1 = 0
    force2 = 0
    force3 = 0
    force4 = 0
    ri1 = coordinates[i]
    ri2 = coordinates[i+1]
    ri3 = coordinates[i+2]
    ri4 = coordinates[i+3]
    for j = 1 to N skip 4
        rj1 = coordinates[j]
        rj2 = coordinates[j+1]
        rj3 = coordinates[j+2]
        rj4 = coordinates[j+3]
             
        force1 += force_function4( ri1, rj1, rj2, rj3, rj4 )
        force2 += force_function4( ri2, rj1, rj2, rj3, rj4 )
        force3 += force_function4( ri3, rj1, rj2, rj3, rj4 )
        force4 += force_function4( ri4, rj1, rj2, rj3, rj4 )
    end
    return force1, force2, force3, force4
end kernel

forces1, forces2, forces3, forces4 = kforce4x4( indices )

\end{verbatim}
\end{codepar}
In the above code, the input is the sequence of integers $1,5,9,...N$
and the output is 4 force streams. \texttt{force\_function4} uses
the 4-way SIMD math available on the GPU to compute 4 forces at a
time. The four output streams can be trivially merged into a single one if
needed.  Results for this kernel can be seen in Figure~\ref{fig:all}.  Once more
the kernel has become instruction-rate limited and its bandwidth is half that of
the maximum bandwidth of the ATI board, but the overall performance has increased
significantly.

\subsection{Optimization for small systems}
\label{sec:small systems}
In all cases, performance is severely limited when the number of
particles is less than about 4000. This is due to a combination of fixed
overhead in executing kernels and the lack of sufficiently many
parallel threads of execution. It is sometimes necessary to process
small systems or subsystems of particles ($N \approx 100-1000$).

For example, in molecular dynamics where forces tend to be short-range in
nature, it is more common to use $O(N)$ methods by neglecting or
approximating the interactions beyond a certain cutoff
distance. However, when using continuum solvent models, the number of particles
is small enough~($N \approx 1000$) that the $O(N^2)$ method is comparable in
complexity while giving greater accuracy than $O(N)$ methods. 

It is common in stellar dynamics to parallelize the individual time step scheme
by using the block time step method~\cite{Block-step}. In this method forces are
calculated on only a subset of the particles at any one time. In some simulations
a small core can form such that the smallest subset might have less than 1000 particles in it.
To take maximal advantage of GPUs it is therefore important to get good performance for
small output stream sizes.

To do this, we can increase the number of parallel threads by decreasing the
j-loop length. For example, the input stream can be replicated twice, with the
j-loop looping over the first $N/2$ particles for the first half of the
replicated stream and looping over the second $N/2$ particles for the second
half of the stream.  Consider the following pseudocode that replicates the
stream size by a factor of 2: 

\begin{codepar} 
\begin{verbatim}

stream coordinates;
stream indices = range( 1 to 2N );
stream partial_forces;

kernel kforce( i )
    force = 0
    if i <= N:
         ri = coordinates[i]
         for j = 1 to N/2
                rj = coordinates[j]
                force = force + force_function( ri, rj )
         end
    else
         ri = coordinates[i-N+1]
         for j = N/2+1 to N
                rj = coordinates[j]
                force = force + force_function( ri, rj )
         end
    endif

    return force
end kernel

partial_forces = kforce( indices )

\end{verbatim}
\end{codepar}

\begin{figure}
	\centering
	\includegraphics[width=2.5in]{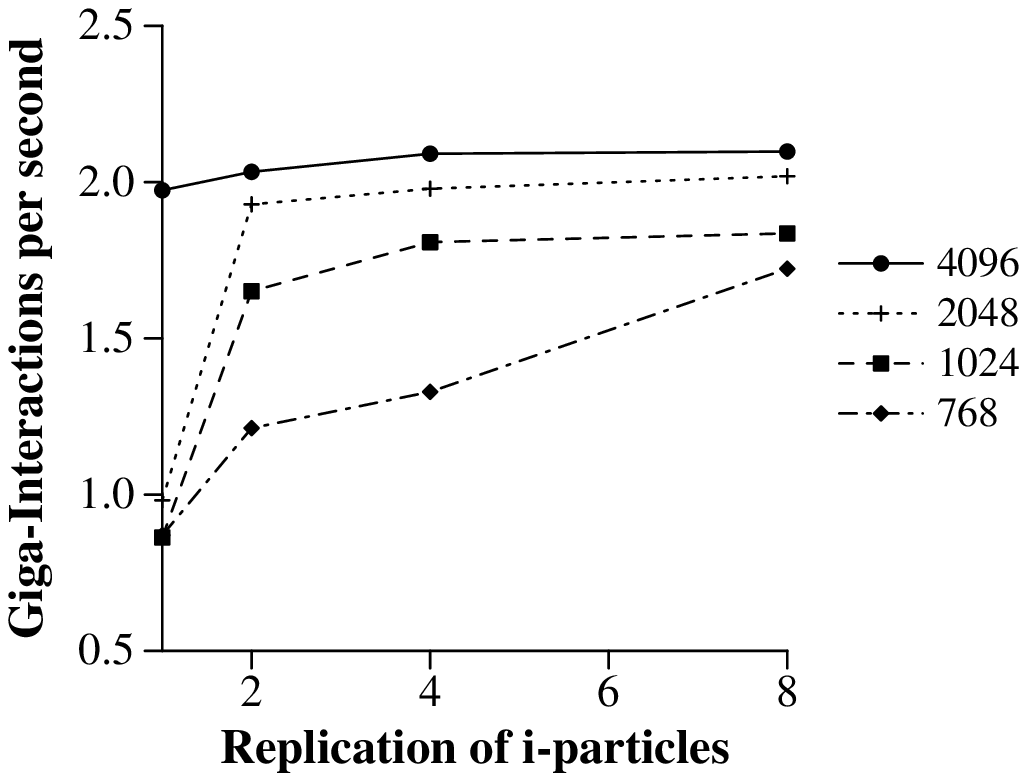}
	\caption{Performance improvement for LJC(sigmoidal) kernel with i-particle replication for several values of $N$}
	\label{fig:replication}
\end{figure}

In this example, the stream \texttt{indices} is twice as long as the
\texttt{coordinates} stream and contains integers in sequence from $1$ to $2N$.
After applying the kernel \texttt{kforce} on \texttt{indices} to get
\texttt{partial\_forces}, the force on particle $i$ can be obtained with by
adding \texttt{partial\_forces[i]} and \texttt{partial\_forces[i+N]}, which can
be expressed as a trivial kernel. The performance of the LJC(sigmoidal) kernel
for different number of replications of the i-particles is shown in Figure~\ref{fig:replication} for several system sizes.

\section {Results}
\label{sec:results}

All kernels were run on an ATI X1900XTX PCIe graphics card on Dell
Dimension 8400 with ATI Catalyst version 7.2 drivers and the
latest DirectX SDK (December 2006). A number of different force models
were implemented with varying compute-to-bandwidth ratios (see
Table~\ref{table:kernels}). A sample code listing is provided in the appendix
(\ref{sec:appendix}) to show the details of how flops are counted. 

To compare against the CPU, a specially optimized version of the
GA and GAJ kernels were written since no software suitable for a direct
comparison to the GPU existed.  The work of~\cite{Grav-SSE} uses SSE for 
the GAJ kernel but
does some parts of the calculation in double precision which makes it
unsuitable for a direct comparison.  The performance they achieved
is comparable to the performance achieved here.
Using SSE intrinsics and Intel's C++ Compiler v9.0, we
were able to obtain sustained performance of 3.8 GFlops on a 3.0 GHz
Pentium 4.

GROMACS~\cite{GROMACS} is currently the fastest performing molecular
dynamics software with hand-written SSE assembly loops.  As mentioned
in Section~\ref{sec:algorithm} the CPU can do out-of-order writes
without a significant penalty. GROMACS uses this fact to halve the
number of calculations needed in each force calculation step. In the
comparison against the GPU in Table \ref{table:gmxgpu} the
interactions per second as reported by GROMACS have been doubled to
reflect this. In MD it is common to use neighborlists to reduce the
order of the force computation to $O(N)$.  The performance of GROMACS
doing an $O(N^2)$ calculation as well as an $O(N)$ calculation for a
80 residue protein (lambda repressor, 1280 atoms) is shown in
Table\ref{table:gmxgpu}.  Despite using a fairly modest cutoff length
of 1.2 nm for the $O(N)$ calculation, the $O(N^2)$ GPU calculation represents
an order-of-magnitude performance improvement over existing methods on CPUs.

\begin{table*}
\centering
\scriptsize
\begin{tabular}{lccccc}

\toprule
              & GMX Million          & GMX $O(N^2)$ & GMX $O(N)$ &    GPU  Million          & GPU    \\
Kernel        & Interactions/sec     & ns/day      &   ns/day    &  Interactions/sec          & ns/day \\

\midrule
LJC(constant) &  66  &   5.6         &   18.2     &   1327      & 140    \\
LJC(linear)*  &  33  &   2.06        &   9.08     &   1327      & 140    \\
LJC(sigmoidal)&  40  &   2.5         &   11       &   1203      & 127    \\
\bottomrule

\end{tabular}
\normalsize
\caption{Comparison of GROMACS(GMX) running on a 3.2 GHz Pentium 4 vs. the GPU
showing the simulation time per day for an 80 residue protein 
(lambda repressor)  *GROMACS does not have an SSE inner loop for LJC(linear) }
\label{table:gmxgpu}
\end{table*}

\section{Discussion}
\subsection{Comparison to other Architectures}
\begin{figure}
   \centering
      	\includegraphics[width=2.5in]{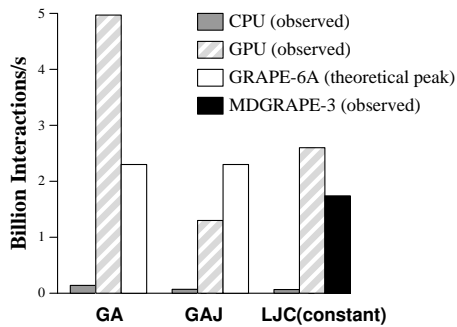}
      	\caption{Speed comparison of CPU, GPU, GRAPE-6A and MDGRAPE-3}
	\label{fig:speed}
\end{figure}

In Figure~\ref{fig:speed} is a comparison of interactions/sec between the ATI X1900XTX, 
GRAPE-6A and a Pentium 4 3.0GHz.  The numbers for the GPU and CPU are observed values,
those for GRAPE-6A are for its theoretical peak. Compared to GRAPE-6A, the GPU can
calculate over twice as many interactions when only the acceleration is computed, and
a little over half as many when both the acceleration and jerk are computed.  The GPU 
bests the CPU by 35x, 39x and 15x for the GA, LJC(constant) and GAJ kernels respectively.

\begin{figure}
   \centering
	\includegraphics[width=2.75in]{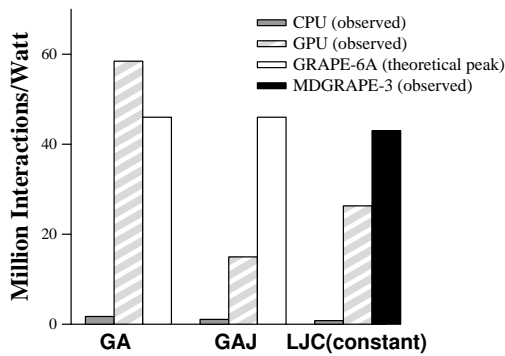}
	\caption{Millions of Interactions per Watt of CPU, GPU, GRAPE-6A and MDGRAPE-3}
	\label{fig:power}
\end{figure}

Another important metric is performance per unit of power dissipated. These results can be seen in
Figure~\ref{fig:power}.  Here the custom design and much smaller on-board memory allows
GRAPE-6A to better the GPU by a factor of 4 for the GAJ kernel, although they are still
about equal for the GA kernel.  The power dissipation of the Intel Pentium 4 3.0
GHz is 82W~\cite{Intel-power}, the X1900XTX is measured to be 85W, we estimate GRAPE-6A's dissipation to be 48W since each of the 
4 processing chips on the board dissipates approximately
12W~\cite{GRAPE_power} and MDGRAPE-3's (MD3-PCIX) dissipation is 40W~\cite{MDGRAPE_3}.

\begin{figure}
   \centering
      	\includegraphics[width=2.5in]{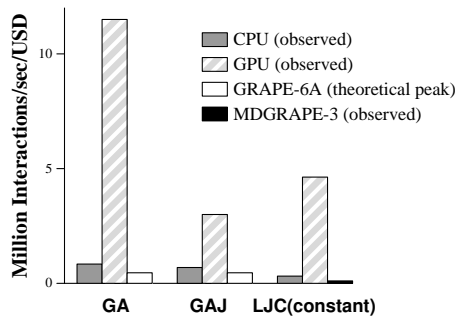}
      	\caption{Useful MFlops per second per U.S. Dollar of CPU, GPU, GRAPE-6A and MDGRAPE-3}
	\label{fig:price}
\end{figure}

The advantages of the GPU become readily apparent when the metric of performance per
dollar is examined (Figure~\ref{fig:price}).  The current price of an Intel Pentium 4
630 3.0GHz is \$100, an ATI X1900XTX is \$350, and an MDGRAPE-3 board costs \$16000~\cite{MDGRAPE_3}.
The GPU outperforms GRAPE-6A by a factor of 22 for the GA kernel and 6 for the GAJ kernel.


\subsection{Hardware Constraints}
The 4$\times$4 unrolling that is possible with the GA kernel does not work
for the other, more complicated kernels.  For example, the GAJ kernel
requires two outputs per particle (jerk in addition to acceleration).
This reduces the maximum unrolling possibility to 2$\times$4 because the GPU
is limited to a maximum of 4 outputs per kernel.  However, even this
amount of unrolling doesn't work because the compiler cannot fit the
kernel within the 32 available registers.  The number of registers is
also what prevents the LJC kernels from being unrolled by 4$\times$4 instead
of 2$\times$4.

This apparent limitation due to the number of registers appears to result
from compiler inefficiencies; the authors are currently hand coding a 2$\times$4
GAJ kernel directly in pixel shader assembly which should cause the
kernel to become compute bound and greatly increase its performance.
The performance gain of unrolling the LJC kernels to 4$\times$4 by rewriting them in 
assembly would most likely be small since these kernels are already compute bound.

While the maximum texture size of 4096$\times$4096 and 512 MB 
would make it possible to store up to 16 million particles on the board
at a time, this really isn't necessary.  In fact, GRAPE-6A only has
storage for 131,000 particles on the board at any one time.  This is
small enough to occasionally seem restrictive - a good balance is
around 1 million particles which could easily be accommodated by 64MB.
If board manufacturers wanted to produce cheaper boards specifically
for use in these kinds of computations they could significantly reduce
the cost without affecting the functionality by reducing the amount of
onboard RAM.

The current limits on the number of instructions also impacts the
efficiency of large GPGPU programs.  On ATI hardware, the maximum shader
length of 512
instructions limits the amount of loop unrolling and
the complexity of the force functions we can handle.  

\subsection{On-board Memory vs. Cache Usage}

As mentioned in Section~\ref{sec:General Optimization} we expect the kernels to make
very efficient use of the cache on the boards.  There are a maximum of 512 
threads in flight on the ATI X1900XTX at any one time~\cite{RADEONX1900}, and in the ideal situation,
each of these threads will try and access the same j-particle at approximately
the same time.  The first thread to request a j-particle will miss the cache and
cause the particle to be fetched from on-board memory, however once it is in the
cache, all the threads should be able to read it without it having to be fetched
from on-board memory again.

For example, in the case of the GA kernel with 65,536 particles, there would be
16,384 fragments to be processed, and if fragments were processed in perfectly
separate groups of 512, then 32 groups would need to be processed.  Each group
would need to bring in 65,536 particles from main memory to the cache resulting
in an extremely low memory bandwidth requirement of 38.2 MB/sec.

\begin{figure}
 \centering
	\includegraphics[width=3.0in]{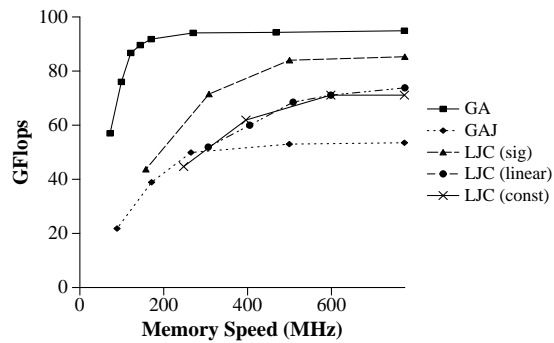}
	\caption{GFlops achieved as a function of memory speed}
	\label{fig:memory}
\end{figure}

Of course, the reality is that particles are not processed in perfectly separate
groups of 512 particles that all request the same particle at the same time, but
by using ATITool~\cite{ATITOOL} to adjust the memory clock of the board we can determine how
much bandwidth each kernel actually needs to main memory.  The results of this
testing can be seen in Figure \ref{fig:memory}.

The performance degradation occurs at approximately 11.3, 5.2, and 2.1
GB/sec for the LJC, GAJ and GA kernels respectively.  The LJC kernels
must also read in an exclusion list for each particle which does not
cache as well as the other reads, and is the reason why their
bandwidth to main memory is higher than that of the gravity kernels.
The number for the GA kernel suggests that approximately 10 particles
are accessing the same j-particle at once.

At memory speeds above 500MHz all the kernels run very near their peak speed, thus
board manufacturers could not only use less RAM, they could also use
cheaper RAM if they were to produce a number of boards that would only be
used for these calculations.  This would reduce the cost and power requirements over the standard high end versions used for gaming.

\subsection{Distributed Computation}
Most biological phenomena of interest occur on timescales currently
beyond the reach of MD simulations. For example, the simplest proteins
fold on a timescale of 5 to 20 microseconds, while more complex
proteins may take milliseconds to seconds. MD simulations on current
generation CPUs are usually limited to simulating about 10 nanoseconds
per day - it would take several years to obtain a 10 microsecond
simulation. However, with the speed increases afforded by the
algorithms and hardware discussed here, we are now be able to simulate
protein dynamics with individual trajectories on the 10 microsecond
timescale in under three months. This will allow the direct simulation
of the folding of fast-folding proteins. Moreover, by incorporating
this methodology into a distributed computing
framework, we are now situated to build Markovian State Models to
simulate even longer timescales, likely approaching
seconds~\cite{MSM-villin}. Thus with the combined effort of GPUs and
distributed computing, one would be able to reach timescales for
folding of essentially all single-domain, two-state folding proteins.
Compared to the donations of CPUs from over 150,000 Windows computers currently producing 145
TFlops, we have 550 GPUs donated to the project producing over 34
TFlops.  Thus each GPU is
providing roughly 60 times the performance of the average donated x86 CPU.

\section{Conclusion}

We have successfully taken advantage of the processing power available
on GPUs to accelerate pairwise force calculations for several commonly
used force models in stellar and molecular dynamics simulations. In
some cases the GPU is more than 25 times as fast as a highly optimized
SSE-based CPU implementation and exceeds the performance of custom
processors specifically designed for these tasks such as GRAPE-6A.
Furthermore, our
performance is compute bound, so we are well poised to take advantage
of further increases in the number of ALUs on GPUs, even if memory
subsystem speeds do not increase significantly. Because GPUs are mass
produced, they are relatively inexpensive and their performance to
cost ratio is an order of magnitude better than the alternatives. The
wide availability of GPUs will allow distributed computing initiatives
like \fah\ to utilize the combined processing power of tens of
thousands of GPUs to address problems in structural biology that were
hitherto computationally infeasible. We believe that the future will
see some truly exciting applications of GPUs to scientific computing.

\appendix

\section{Appendix}
\subsection{Flops Accounting}
\label{sec:appendix}
To detail how we count flops we present a snippet of the actual Brook code for the GA kernel.
The calculation of the acceleration on the first i-particle has been commented with our
flop counts for each instruction.  In total, the calculation of the acceleration on the 
first i-particle performs 76 flops.  Since four interactions are computed, this
amounts to 19 flops per interaction.

\begin{codepar}
\scriptsize
\begin{verbatim}

float3 d1, d2, d3, d4, outaccel1;
float4 jmass, r, rinv, rinvcubed, scalar;

d1 = jpos1 - ipos1; //3
d2 = jpos2 - ipos1; //3
d3 = jpos3 - ipos1; //3
d4 = jpos4 - ipos1; //3

r.x = dot( d1, d1 ) + eps; //6
r.y = dot( d2, d2 ) + eps; //6
r.z = dot( d3, d3 ) + eps; //6
r.w = dot( d4, d4 ) + eps; //6

rinv = rsqrt( r );           //4
rinvcubed = rinv*rinv*rinv;  //8
scalar = jmass * rinvcubed;  //4
outaccel1 += scalar.y * d2 + scalar.z * d3 + scalar.w * d4; //18

if ( Ilist.x != Jlist1.x ){    //don't add force due to ourself
   outaccel1 += scalar.x * d1; //6
}

\end{verbatim}
\normalsize
\end{codepar}

\bibliographystyle{acmsiggraph}
\begin{spacing}{0.05}
\bibliography{nbody}
\end{spacing}

\end{document}